\documentclass[aps,pra,twocolumn,superscriptaddress]{revtex4-1}
\usepackage{graphicx,enumerate,verbatim,bbold}
\usepackage{amsmath,amssymb,amsthm,float,mathrsfs}
\usepackage{dsfont}
\usepackage[colorlinks]{hyperref}
\usepackage[T1]{fontenc}
\usepackage[utf8]{inputenc}
\usepackage{lmodern}
\usepackage{braket}
\usepackage[caption=false]{subfig}
\usepackage{dcolumn}

\usepackage[normalem]{ulem}

\usepackage[usenames,dvipsnames]{color}

\begin{document}

\title{Extended Hubbard model for mesoscopic transport in donor arrays in silicon}

\author{Nguyen H. Le}
\affiliation{Advanced Technology Institute and Department of Physics, University of Surrey, Guildford GU2 7XH, United Kingdom}

\author{Andrew J. Fisher}
\affiliation{UCL Department of Physics and Astronomy and London Centre for Nanotechnology, University College London, Gower Street, London WC1E 6BT, United Kingdom}

\author{Eran Ginossar}
\affiliation{Advanced Technology Institute and Department of Physics, University of Surrey, Guildford GU2 7XH, United Kingdom}

\begin{abstract}
Arrays of dopants in silicon are promising platforms for the quantum simulation of the Fermi-Hubbard model. We show that the simplest model with only on-site interaction is insufficient to describe the physics of an array of phosphorous donors in silicon due to the strong intersite interaction in the system. We also study the resonant tunneling transport in the array at low temperature as a mean of probing the features of the Hubbard physics, such as the Hubbard bands and the Mott gap. Two mechanisms of localization which suppresses transport in the array are investigated: The first arises from the electron-ion core attraction and is significant at low filling; the second is due to the sharp oscillation in the tunnel coupling caused by the intervalley interference of the donor electron's wavefunction. This disorder in the tunnel coupling leads to a steep exponential decay of conductance with channel length in one-dimensional arrays, but its effect is less prominent in two-dimensional ones. Hence, it is possible to observe resonant tunneling transport in a relatively large array in two dimensions. 
\end{abstract} 


\maketitle

\section{Introduction}
Advanced experimental techniques such as single-ion implantation and scanning tunneling microscope (STM) lithography \cite{clark_progress_2003,zwanenburg_silicon_2013} have enabled the fabrication of dopants in silicon with nanometer precision. This opens the prospect of engineering an array of dopants in any desired lattice. Analogous to atoms in optical lattices, these artificial atoms have the potential to be a good platform for simulating quantum many-body physics. Compared with cold-atom simulators \cite{bloch_quantum_2012}, arrays of dopants offer access to systems with stronger correlations, longer-range interactions and a better possibility for realizing the zero-temperature limit \cite{salfi_quantum_2016}. Another distinctive feature of this dopant-based quantum simulator is the availability of transport measurements for probing the relevant properties of the underlying many-body physics.

One of the most important models of strongly-correlated electrons is the Hubbard model. Despite its simplicity, the Hubbard model and its variants are believed to cover a wide range of exciting phenomena such as unconventional superconductivity \cite{gull_superconductivity_2013}, quantum spin liquids \cite{balents_spin_2010}, and Nagaoka ferromagnetism \cite{nielsen_nanoscale_2007,tasaki_nagaokas_1998}. Many of these have not been fully understood owing to the lack of reliable numerical and analytical solutions in two dimensions. This has spurred many proposals and experimental realizations of the Hubbard model with analog quantum simulators based on cold atoms \cite{hart_observation_2015}, quantum dot arrays \cite{hensgens_quantum_2017, byrnes_quantum_2008} and dopants in silicon \cite{salfi_quantum_2016}.

Much of the initial work on the physics of dopant arrays has been based on the assumption that the Hubbard model with only on-site interactions is a good effective Hamiltonian for describing the physics of the system. A few examples are the proof-of-principle experiment on simulating the Hubbard model with two boron acceptors \cite{salfi_quantum_2016}, and transport measurements in the highly disordered one dimensional (1D) chain of phosphorous donors in silicon \cite{prati_anderson-mott_2012,prati_band_2016}. In the first part of this paper we ask whether the above assumption is justified. We find that models with only on-site interactions are not sufficient when the system is far from half-filling, as inter-site Coulomb interactions become important. Thus, it is more accurate to say that the donor array simulates the extended version of the Hubbard model where the long range interaction is included.

\begin{figure}[t]
\includegraphics[scale=0.45]{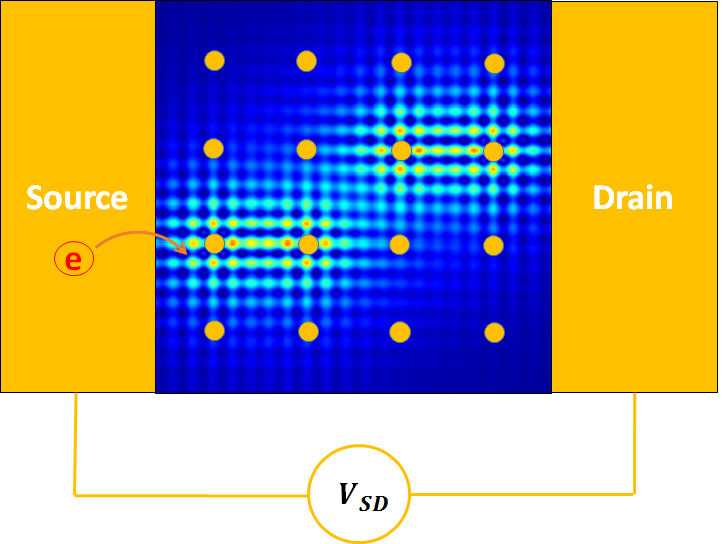}
\caption{A 2D donor array coupled to a source and a drain under a bias $V_{\text{SD}}$. The chemical potential of the leads can be varied by a gate voltage. Electrons from the source can tunnel through a many-body state of the array that is delocalized along a path that connects one side of the array to the other.}\label{fig:array}
\end{figure}

Transport measurements are likely to be one of the most useful probes for a dopant-based quantum simulator. In the second part of the paper we study the resonant tunneling transport in finite arrays of donors connected to two leads (see Fig.~\ref{fig:array}) using  exact diagonalization for small sizes and approximation methods including Hartree-Fock mean field theory for large sizes. We obtain the resonant conductance spectrum at low temperature. The long range repulsion between the electrons leads to broadening of the lower and upper Hubbard bands, which is reflected in the conductance spectrum. Moreover, long range electron-ion core attraction localizes the electrons towards the center of the array at low filling, causing a suppression of transport.

Another important effect that we study is the sharp oscillation in the tunnel coupling due to the intervalley interference of the donor  electron's wavefunction \cite{koiller_exchange_2001,hu_charge_2005,wellard_donor_2005,gamble_multivalley_2015}. This is known to complicate a precise implementation of the Kane's silicon-based quantum computer \cite{kane_silicon-based_1998} and is likely to pose similar challenges for the development of a quantum simulator. Even when the donors are placed with only nanometer uncertainty, the tunnel coupling oscillation results in strong disorder and hence the localization of the charge excitation responsible for transport. We find that this disorder suppresses enormously the transport at low temperature in one-dimensional arrays. The situation for two-dimensional (2D) arrays is more promising, owing to the larger number of possible paths for transport. Our simulation shows that it is possible to observe the resonant tunneling current in a relatively large 2D array (with size up to $10 \times 10$) in spite of the strong disorder in the tunnel coupling. We obtain numerically the scaling of the conductance with array size in both 1D and 2D.

In the final sections we discuss potential experimental deviations from our theoretical calculation and offer a summary of the main results in the conclusions. Technical details and the computational codes used in this paper are provided in the Appendix.

\section{The extended Hubbard Hamiltonian}
We first derive the appropriate extended Hubbard model with intersite interactions to describe the physics of the donor array. In effective-mass theory, the ground state of an electron bound to an isolated donor in silicon is the $\text{1sA}_1$ multivalley coupled wavefunction  \cite{kohn_theory_1955}
\begin{equation}\label{eq:gs}
\psi(\mathbf{r})=\sum_{\mu} F_{\mu}(\mathbf{r})\phi_{\mu}(\mathbf{r}),
\end{equation}
where $\mu=x,-x,y,-y,z,-z,$  indicates the six conduction band minima of silicon, $\phi_{\mu}(\mathbf{r})=e^{i \mathbf{k_{\mu}} \cdot \mathbf{r}} u_{\mu}(\mathbf{r})$ is the Bloch function and $F_{\mu}(\mathbf{r})$ the envelope function for each valley.

If the distance between adjacent donors in an array is sufficiently small, the overlap of the ground states between nearest neighbors leads to a formation of an energy band. This band can be described in second quantization by the extended Hubbard model \cite{DionysBaeriswyl826}
\begin{align}\label{eq:HubbardH}
H_{\text{array}}=&\sum_{i }\epsilon_{i} n_{i }-\sum_{\braket{i j} } \sum_{\sigma=\uparrow,\downarrow} t_{ij}\left(c^{\dagger}_{i \sigma} c_{j \sigma} +c^{\dagger}_{j \sigma} c_{i \sigma}\right)  \\ \nonumber
&+ U\sum_{i} n_{i\uparrow} n_{i \downarrow} +\sum_{i\neq j} W_{ij} n_i n_j,
\end{align}
where {$\epsilon_{i}$ is the single site energy at site $i$, $t_{i j}$ the tunnel coupling between site $i$ and site $j$, $c^{\dagger}_{i \sigma}$ the creation operator of an electron with spin $\sigma$ at site $i$,  and $U$ the on-site interaction. The necessity of including the long range electron-electron repulsion $W_{i j}$ due to its large strength is discussed below. The sum over $\braket{i j}$ is understood to be over nearest-neighbor pairs only; this simplification in the kinetic term is justified by the fact that the tunnel coupling $t_{ij}$ between two donors decays exponentially with the donor separation.  A full configuration interaction study of a pair of neutral donors for the 1s manifolds in Ref.~\cite{saraiva_theory_2015} shows that the Hubbard approximation is sufficient for describing ground state properties when the donor separation is a few times larger than the scaled Bohr radius ($d\gtrsim 5$ nm for Si:P).

At each donor site the energy of a single electron (measured with respect to the conduction band minimum) is the ground-state energy of the $\text{1sA}_1$ level perturbed by the long-range Coulomb attraction from the ion cores of all the other donors in the array; $\epsilon_i=-E_B+\sum_{j\neq i} V_{i j}$ where $E_B\approx 45$ meV is the binding energy of an isolated neutral $D^{0}$ center and 
\begin{align}\label{eq:Vij}
V_{i j}=-V_0 \int \frac{ |\psi(\mathbf{r}-\mathbf{R}_i)|^2}{|\mathbf{r}-\mathbf{R}_j|} \mathrm{d} \mathbf{r}
\end{align}
is the long range Coulomb attraction from the ion core at site $j$. Here $V_0=e^2/(4\pi \epsilon_0 \epsilon_{Si})\approx 123$ meV$\times$nm with $\epsilon_{Si}=11.6$. 

The tunnel coupling $t$ between two phosphorous donors in silicon has been previously studied in the context of  silicon-based quantum computer architecture \cite{hu_charge_2005}. We follow the same approach of estimating $t$ as half the energy separation between the symmetric and antisymmetric linear combination of the $\text{1sA}_1$ wavefunctions at the two donor sites, which yields
\begin{align}\label{eq:tij}
t_{ij}=\frac{S_{ij}V_{ij}-V'_{ij}}{1-S_{ij}^2},
\end{align}
where 
\begin{align}
S_{i j}=\int \psi^*(\mathbf{r}-\mathbf{R}_i)\psi(\mathbf{r}-\mathbf{R}_j) \mathrm{d} \mathbf{r}
\end{align}
is the overlap,
\begin{align}
V'_{i j}=-V_0 \int \frac{\psi^*(\mathbf{r}-\mathbf{R}_i)\psi(\mathbf{r}-\mathbf{R}_j)}{|\mathbf{r}-\mathbf{R}_j|}\mathrm{d} \mathbf{r},
\end{align}
and $V_{ij}$ is the integral given in Eq.~\eqref{eq:Vij}.}

The on-site interaction between two electrons bound to the same donor can be obtained from the binding energy of the negatively charged $D^{-}$ center; $U=E_B(D^{0})-E_B(D^{-})\approx 43$ meV. The intersite electron-electron repulsion is given by
\begin{align}\label{eq:Wjk}
W_{ij}=&V_0 \int 
\frac{|\psi(\mathbf{r}_1-\mathbf{R}_i)|^2 |\psi(\mathbf{r}_2-\mathbf{R}_j)|^2 }{|\mathbf{r}_1-\mathbf{r}_2|} \mathrm{d} \mathbf{r_1} \mathrm{d} \mathbf{r_2}.
\end{align}
\begin{figure}[t]
\includegraphics[scale=0.4]{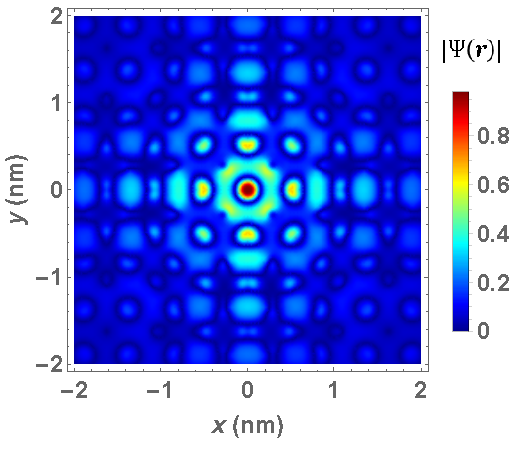}
\caption{Density plot of the central cell-corrected multivalley ground state wavefunction in the (100) plane. This wavefunction is more sharply peaked at the donor's site compared with the Kohn-Luttinger wavefunction \cite{kohn_theory_1955} due to the strong attraction of the central cell potential. The oscillation in density is due to the Bloch part of the wavefunction.}\label{fig:1sA}
\end{figure}

To evaluate these integrals we use the full $\text{1sA}_1$ wavefunction of Ref.~\cite{gamble_multivalley_2015}, depicted in Fig.~\ref{fig:1sA}. This wavefunction includes the periodic part $u(\mathbf{r})$ of the Bloch functions, calculated by density functional theory, and the envelope function $F(\mathbf{r})$ obtained from Shindo-Nara multivalley effective mass theory \cite{shindo_effective_1976}. As this wavefunction shows a good quantitative agreement with the results observed in STM measurements \cite{salfi_spatially_2014}, we expect that it also gives a reliable estimate for the parameters of the effective Hubbard Hamiltonian. The integrands are highly oscillatory due to the Bloch part of $\psi(\mathbf{r})$, but the integrals can be evaluated with satisfactory accuracy  by Monte-Carlo integration with importance sampling. We use the numerical package CUBA \cite{hahn_cubalibrary_2005} for this purpose. 
\begin{figure}[t]
\includegraphics[scale=0.55]{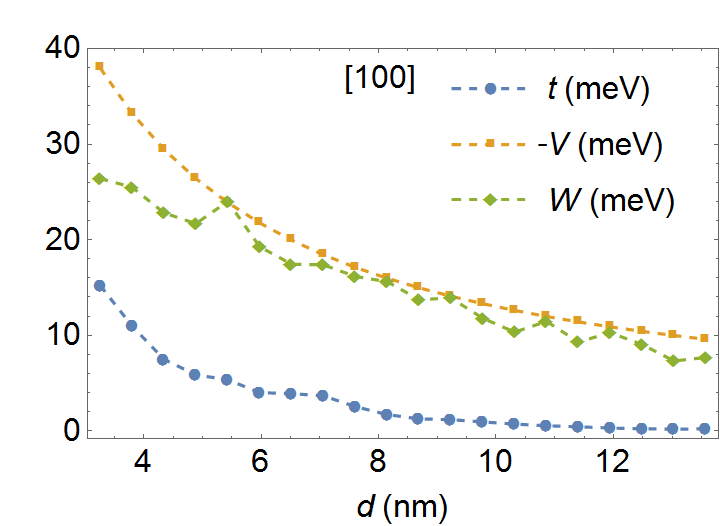}
\\
\includegraphics[scale=0.55]{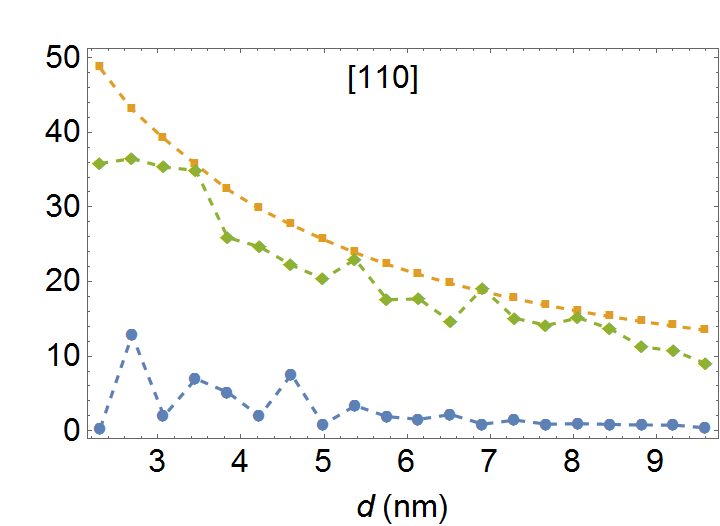}
\\
\includegraphics[scale=0.55]{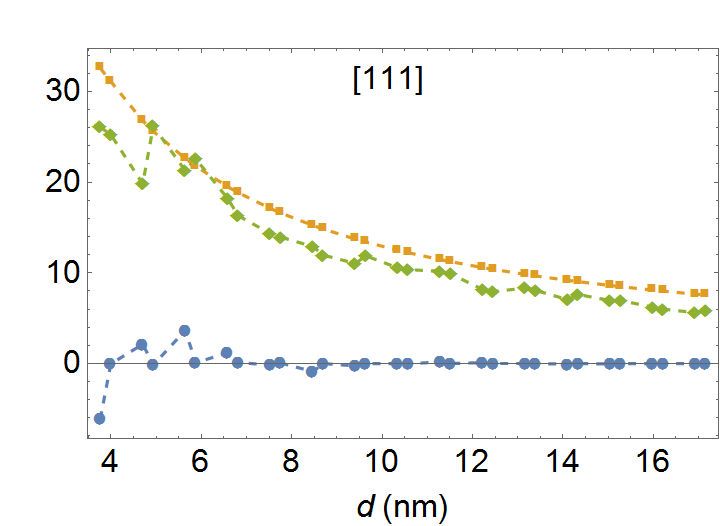}
\caption{Tunnel coupling $t$ and long range interactions $V$ and $W$ for various donor separations along the silicon crystal axes [100], [110], and [111]. The long range interactions are considerably larger than the tunnel coupling for all donor separations.}\label{fig:param}
\end{figure}

Figure \ref{fig:param} shows the tunnel coupling and the long range interactions for the three high-symmetry silicon crystal axes [100], [110], and [111]. The oscillation in the tunnel coupling due to the intervalley interference is most visible for the [110] and [111] directions. A donor array fabricated by STM lithography is most conveniently aligned along the [110] direction of the zigzag silicon bond chain \cite{weber_ohms_2012}, so we focus on this direction for the rest of our paper. It is remarkable that for all three directions the long-range interactions are much larger than the tunnel coupling. This demonstrates the need for including these interactions in the effective Hubbard model for an accurate description of the physics. In our calculation for small arrays we find that it is important to include the long range interaction from \textit{all} the electron-electron and electron-ion core pairs in the array, not only the nearest-neighbor pairs.  For all directions the intersite electron-electron and electron-nuclear interactions approximately cancel at large separations: $V_{ij}\approx -W_{ij}$.

\begin{figure}[t]
\includegraphics[scale=0.5]{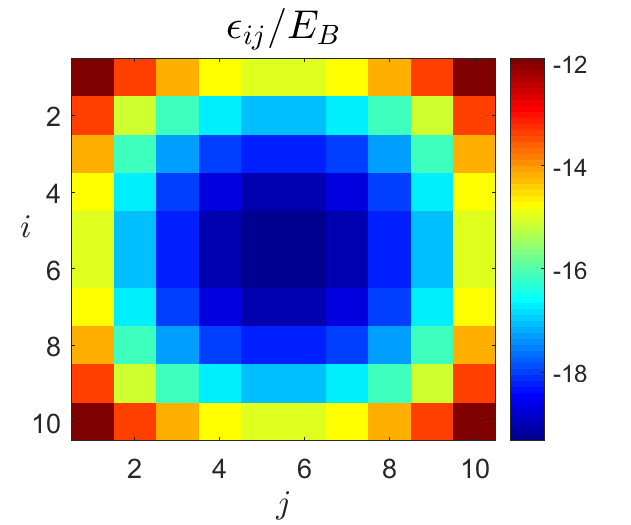}
\caption{A density plot of the single site energy, scaled by the binding energy of an isolated neutral donor, in a $10 \times 10$ array on a square lattice ($i$ and $j$ are the row and column indices of the sites in the array, respectively). The nearest-neighbor separation is $4.6$ nm. There is a huge variation in the single site energy due to the long range attraction from all the ion cores in the array.}\label{fig:sserg}
\end{figure}

An obvious effect of the long range Coulomb attraction between electrons and donor's ion cores $V_{ij}$ is that sites at the edge of the array have higher energy than those at the center. An illustration of the spread in $\epsilon_i$ is shown in Fig.~\ref{fig:sserg}. At low electron filling this spread may lead to the strong localization of the wavefunction towards the center of the array. This is discussed in more details in Sec.~\ref{sec:adderg}.

We note that the most general form of the many-electron Hamiltonian for the array in second quantization includes long range tunneling terms and interacting terms of the form $V_{ijkl}c^{\dag}_i c^{\dag}_j c_k c_l$ of which the on-site interaction term and the intersite Coulomb repulsion term are special cases \cite{FabianH.L.Essler719}. In our calculation we verify that, for the range of parameters realized with donor arrays, the long range tunneling terms and the extra interaction terms have an insignificant effect on the ground state properties of an ordered array, as well as the order of magnitude of the conductance of a disordered array discussed in Sec.~\ref{sec:condctdist}.

\section{Resonant tunneling transport}
One of the advantage of quantum simulation with an array of donors is the available access to transport measurements. In this section we study how a measurement of the conductance can reveal the spectral features of the Hubbard physics in the array. Such an experiment can also serve as a verification of the validity of the Hubbard approximation for the array.

There have been a few experiments on the transport in a 1D chain of phosphorous donors in silicon, fabricated by ion implantation \cite{prati_band_2016, prati_anderson-mott_2012}. In these devices there are large uncertainties in the position of the donor and hence the transport mechanism is mainly phonon-assisted hopping between localized states, which is different from the resonant tunneling through delocalized states considered here. Moreover, the effect of long range interaction is often neglected in the analysis of the previously done experiments \cite{prati_band_2016, prati_anderson-mott_2012,salfi_quantum_2016}. We show that while this may be justified at half filling, long range interaction must be taken into account at lower filling.

\subsection{Addition energy: lower and upper Hubbard bands}\label{sec:adderg}
We consider a small 2D array coupled to the source and drain as shown in Fig.~\ref{fig:array}. We suppose the chemical potential $\mu$ in the leads can be varied by a gate voltage. We are interested in the conductance of the array in the sequential tunneling regime at low temperature. 
\begin{figure}[t]
\includegraphics[scale=0.6]{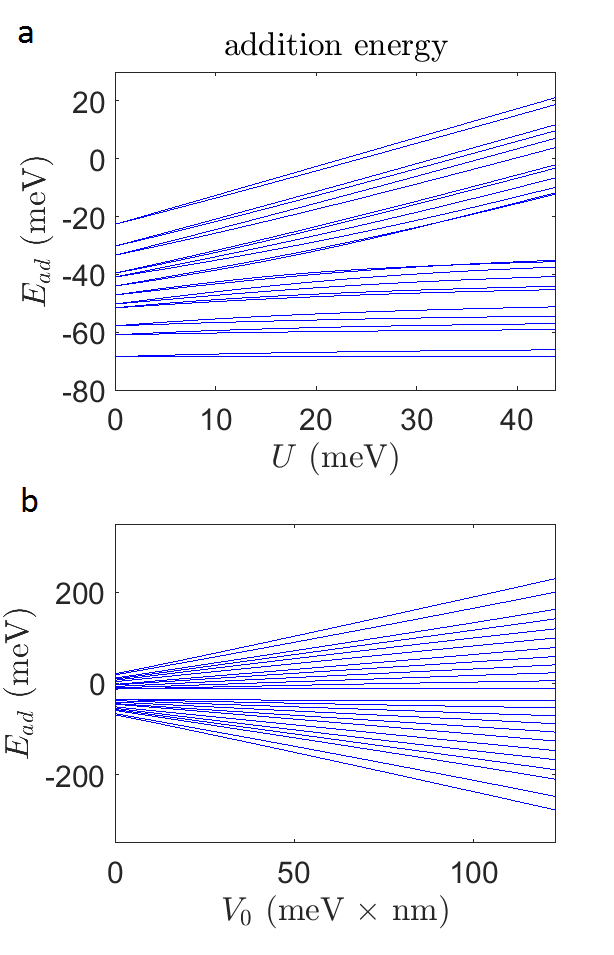}
\caption{Addition energy spectrum for a $3\times 4$ donor array as (a) on-site interaction is increased to its true value of 43.86 meV while long range interaction is neglected; and (b) with on-site interaction at $43.86$ meV and long-range interaction increased to its true value of 123 meV$\times$nm. The on-site interaction opens the Mott gap; and the intersite interaction leads to broadening of the lower and upper Hubbard bands.}\label{fig:eadd}
\end{figure}

The Hilbert space for the eigenstates of the effective Hamiltonian $H_{array}$ can be divided into different sectors characterized by the electron number $n$ which is controlled by the chemical potential . For each sector the few states with lowest energies can be computed by exact diagonalization with the Lanczos algorithm \cite{HolgerFehske789}, which can be sped up with the use of graphic processing units \cite{siro_exact_2012}. 

The resonant tunneling can be described by the rate equation developed for a single quantum dot \cite{beenakker_theory_1991} and later generalized for a 1D chain of quantum dots \cite{chen_resonant_1994}. In this formalism the conductance is computed from the tunneling rate of the electrons from the source to the system and from the system to the drain. It is assumed that the tunneling between the system and the electrodes is incoherent, but that once within the system the electron transport is coherent and elastic; tunneling is therefore possible only when  $\mu = E_m(n)-E_0(n-1)$ where $E_m(n)$ is the energy eigenvalue for state $m$ of the sector with $n$ electrons, and $E_0(n)$ is the lowest eigenenergy. In other words the energy of the electron in the leads ($\mu$) must be large enough to compensate for the increase in the energy of the donor arrays when it tunnels to the array and thus raises the filling from $n-1$ to $n$. Therefore, an important parameter for determining the chemical potential at which resonant transport is allowed is the addition energy $E_{ad}(n)= E_0(n)-E_0(n-1)$. This addition energy reduces to the single-particle energy spectrum in the noninteracting regime; thus it is the many-body analog of single-particle levels.

The Hubbard model can be treated with perturbation theory in the weak interaction regime $U/t\ll 1$. For donor arrays in silicon with donor separation larger than the scaled Bohr radius, $U/t \gg 1$, which is a regime that is not tractable by classical methods \cite{simons_solutions_2015} and hence is the chief target of quantum simulation \cite{salfi_quantum_2016}. We focus on arrays with small donor separation so that the tunneling current through the array is as large as possible. For our study we choose a nearest neighbor separation of $d=4.6$ nm, which is around the minimum separation for which the Hubbard approximation is still valid \cite{saraiva_theory_2015}.

Figure \ref{fig:eadd} shows how the addition energy spectrum depends on interactions for a $3\times 4$ array of donors on a square lattice with a nearest neighbor separation of $d=4.6$ nm. The corresponding tunnel coupling is $t\approx 7.5$ meV. Each donor can host one electron in the $D^{0}$ state or two electrons in the $D{^-}$ state, resulting in a maximum filling of 24 for the array. Each line from bottom to top shows the addition energy for $n=1,\dots, 24$, respectively. In order to demonstrate separately the effects of the on-site interactions and long range interactions on the spectrum, we first plot the spectrum as the on-site interaction is increased from zero to 43.86 meV while long range interaction is neglected in Fig.~\ref{fig:eadd}(a) and then continue with turning on the long range interaction in Fig.~\ref{fig:eadd}(b). To simplify our calculation we use the point-charge values $\mp V_0/|\mathbf{R_i}-\mathbf{R_j}|$ to approximate $V_{ij}$ and $W_{ij}$, which is close to the values of the full integrals for the range of donor separation considered here.

The on-site interaction results in the opening of the Mott gap separating the lower and upper Hubbard bands. In the lower band the number of electrons is less than the number of donor sites so the electrons can avoid each other, explaining the low energy required for adding another electron to the array. When this lower band is fully filled adding electrons requires paying the energy $U$ due to double occupancy on a donor, hence the jump from the lower to the upper band. The reflection symmetry between the two bands around the midpoint of the Mott gap is due to the particle-hole symmetry of the Hubbard model on a square lattice.

\begin{figure}[t]
\includegraphics[scale=0.5]{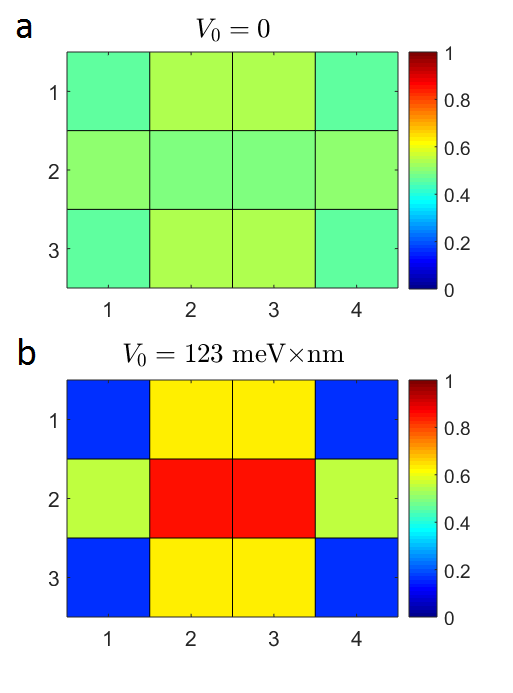}
\caption{Electron number distribution in the array at quarter filling with (a) no long range interactions and (b) full long range interactions.}\label{fig:density}
\end{figure}

The long range interactions lead to the broadening of both the lower and upper bands, which is due to the large  electron-electron interaction $W$ required to add an electron at any given filling. We note that there is little change in the two energy levels around half filling (the top of the lower band and bottom of the upper band), which can be understood by the fact that at half filling the long range electron-electron repulsion is largely canceled by the long range electron-ion core attraction. At other fillings the number of electrons and ion cores are not equal so this cancellation is not perfect, thus the long range interactions have big effects on the energy levels and other properties of the ground state. As an illustration the change in the electron number distribution at one quarter filling is shown in Fig.~\ref{fig:density}: Without long range interactions the electrons are evenly distributed throughout the array; but in reality the electron-ion core attraction, which is dominant at low filling, leads to a localization of the electrons towards the center.  We also confirm with our numerical simulation that there is virtually no change in the electron number distribution when the array is half filled: In this case each site is occupied with one electron regardless of the value of $V_0$.

The reader may notice that in the estimation of the single site energy variation we do not include possible interactions with Coulomb centers outside the array, most importantly the image charges in the source and drain leads. We show in the Appendix that, for a reasonable geometry, although the image charges cause large shifts in the magnitude of the single site energy, there is little change in the energy difference from site to site, which is the relevant physical quantity.

\subsection{Conductance spectrum}\label{sec:gspectrum}
When the array is in close proximity to the source and drain conducting leads electrons can tunnel in and out of the array. This tunneling happens across an interface electrostatic potential barrier which forms between the leads and the array when the electrostatic potentials of the two regions meet \cite{fuechsle_single-atom_2012}. The donors behave as a set of potential wells through which the electrons from the source can tunnel to the drain. The Hamiltonian of the total system including the donor array and the leads is \cite{chen_resonant_1994,klimeck_electron-phonon_2008}. 
\begin{align}
H=H_{\text{array}}+H_{\text{leads}}+H_{\text{coupling}},
\end{align}
where the left (L) and right (R) leads are assumed to have a simple noninteracting single band of energy $\epsilon_k$ at momentum $k$
\begin{align}
H_{\text{leads}}=\sum_{k\sigma}\epsilon^{(R)}_{k\sigma} c^{\dagger}_{k\sigma} c_{k\sigma}+\sum_{k\sigma}\epsilon^{(L)}_{k\sigma} c^{\dagger}_{k\sigma}c_{k\sigma},
\end{align} 
and the coupling is only important between the leads and the donors nearest to them
\begin{align}
H_{\text{coupling}}=\mathcal{V}\bigg[&\sum_{k \sigma,j\in \text{cL}} (c^{\dagger}_{k \sigma}c_{j\sigma}+h.c.)\\ \nonumber
&\sum_{k \sigma,j\in \text{cR}} (c^{\dagger}_{k \sigma}c_{j\sigma}+h.c.)\bigg],
\end{align}
where cL (cR) indicates the leftmost (rightmost) column of the array. A symmetric set-up where the couplings to the right and the left lead are identical is assumed in the above formula. The coupling strength $\mathcal{V}$ depends mainly on the potential barrier at the lead-donor interface and decreases exponentially with the separation between the donors and the leads. 

The eigenstates of $H_{array}$ can be labeled by the quantum numbers $n_{\uparrow}, n_{\downarrow},$ and $\alpha=0,1,2,\dots$ indicating the different eigenstates with the same $n_{\uparrow}, n_{\downarrow}$. Let us denote the many-body wavefunction by $\Psi^{n_{\uparrow}, n_{\downarrow}}_{\alpha}$ and its energy $E^{n_{\uparrow}, n_{\downarrow}}_{\alpha}$. When an electron with spin $\sigma$, momentum $k$ and energy $\epsilon_k$ tunnels from one of the leads to the array the state of the array can change from ($n_{\sigma}-1, n_{\bar{\sigma}}, \beta$) to ($n_{\sigma}, n_{\bar{\sigma}}, \alpha$). Assuming that the process is elastic, resonant tunneling happens when $\epsilon_k = E^{n_{\sigma},n_{\bar{\sigma}}}_{\alpha}-E^{n_{\sigma}-1,n_{\bar{\sigma}}}_{\beta}$. The rate of tunneling from the left lead can be obtained from the Fermi's golden rule \cite{klimeck_electron-phonon_2008}
\begin{align}\label{eq:rate}
\Gamma^{(L),n_{\sigma},n_{\bar{\sigma}}}_{\alpha,\beta,\sigma}&=\Gamma M^{(L),n_{\sigma},n_{\bar{\sigma}}}_{\alpha,\beta,\sigma},
\end{align}
where $\Gamma=2\pi \mathcal{V}^2$ and the matrix element is
\begin{align}
M^{(L),n_{\sigma},n_{\bar{\sigma}}}_{\alpha,\beta,\sigma}=\sum_{j\in cL}  |\braket{\Psi^{n_{\sigma},n_{\bar{\sigma}}}_{\alpha}|c^{\dagger}_{j\sigma}|\Psi^{n_{\sigma}-1,n_{\bar{\sigma}}}_{\beta}}|^2. 
\end{align}
The  rate for tunneling from the right lead has the same expression with L $\leftrightarrow$ R. From the form of the matrix element one can think of the tunneling process as creating a charge excitation in the leftmost column of the array on top of the existing electrons prepared in the state $\Psi^{n_{\sigma}-1,n_{\bar{\sigma}}}_{\beta}$. This matrix element vanishes unless the added charge excitation in the state $\Psi^{n_{\sigma},n_{\bar{\sigma}}}_{\alpha}$ has nonzero probability density at a site in the leftmost column.

Utilizing the above tunneling rate in the rate-equation formalism developed for a quantum dot \cite{beenakker_theory_1991} one arrives at the following expression for the linear response conductance at temperature $T$
\begin{align}\label{eq:condct}
G&=g_T\sum_{n_{\sigma},n_{\bar{\sigma}}}\sum_{\alpha,\beta,\sigma}\frac{M^{(L),n_{\sigma},n_{\bar{\sigma}}}_{\alpha,\beta,\sigma} M^{(R),n_{\sigma},n_{\bar{\sigma}}}_{\alpha,\beta,\sigma}}{M^{(L),n_{\sigma},n_{\bar{\sigma}}}_{\alpha,\beta,\sigma}+ M^{(R),n_{\sigma},n_{\bar{\sigma}}}_{\alpha,\beta,\sigma}}\\ \nonumber
&\times P^{n_{\sigma},n_{\bar{\sigma}}}_{\alpha}\left[1-f_{FD} \left(E^{n_{\sigma},n_{\bar{\sigma}}}_{\alpha}-E^{n_{\sigma}-1,n_{\bar{\sigma}}}_{\beta}-\mu\right)\right],
\end{align}
where $g_T=e^2\Gamma/(\hbar\, kT)$ and
\begin{align}
P^{n_{\sigma},n_{\bar{\sigma}}}_{\alpha}=\frac{\exp\left[-(1/kT)\left(E^{n_{\sigma},n_{\bar{\sigma}}}_{\alpha}-n\mu\right) \right]}{\sum_{n_{\sigma},n_{\bar{\sigma}},\alpha}\exp\left[-(1/kT)\left(E^{n_{\sigma},n_{\bar{\sigma}}}_{\alpha}-n\mu\right) \right]}
\end{align} 
is the grand canonical ensemble probability that the array is in the state $\Psi^{n_{\sigma},n_{\bar{\sigma}}}_{\alpha}$ at equilibrium and $f_{FD}$ the Fermi-Dirac distribution function \cite{chen_resonant_1994}.  The probability $P^{n_{\sigma},n_{\bar{\sigma}}}_{\alpha}$ makes sure that only low-lying energy levels  contribute to the conductance when $kT$ is small compared with the energy separations. Thus, for each sector $\{n_{\uparrow}, n_{\downarrow}\}$ one needs only compute a few eigenstates, saving a lot of computational effort. 

This rate-equation formula for the conductance is valid in the weak coupling regime, that is, when the tunneling rate between the leads and the donors satisfies $\Gamma\ll \Delta E$ where $\Delta E$ is the gap between the low-lying energy levels of the donors array. This has been realized in experiments with dopants in silicon \cite{fuechsle_single-atom_2012,jehl_coupled_2015}, and is also the relevant regime for probing a quantum simulator as the  coupling between the probe and the simulated system should be weak enough so as not to disturb the physics of the system. 

The product $M^{(L),n_{\sigma},n_{\bar{\sigma}}}_{\alpha,\beta} M^{(R),n_{\sigma},n_{\bar{\sigma}}}_{\alpha,\beta}$ in the conductance is nonvanishing only when the added charge excitation in the state $\Psi^{n_{\sigma},n_{\bar{\sigma}}}_{\alpha}$ has nonzero probability density at both the left and right ends of the array, which means this charge excitation must be delocalized over the whole length of the array. A localization of this quasiparticle leads to a vanishing conductance; thus computing the conductance is useful  for studying the degree of localization induced by disorder even in the presence of strong interactions where the picture of a single-particle wavefunction is no longer available.

\begin{figure}[t]
\includegraphics[scale=0.6]{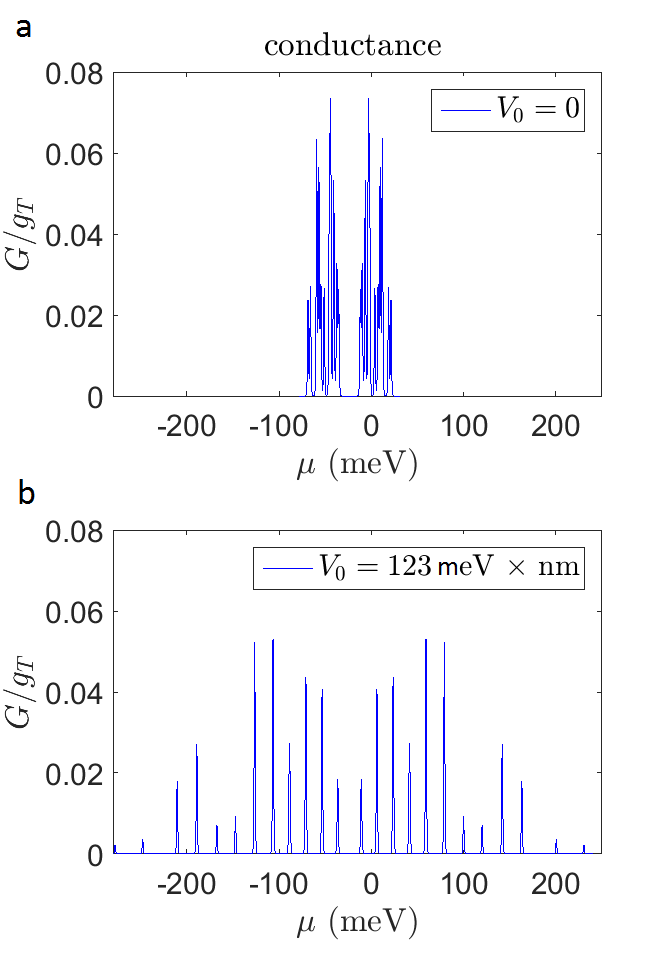}
\caption{(a) Conductance spectrum [scaled by  $g_T=e^2 \Gamma/(\hbar \,kT)$] at 4K for a 3 $\times$ 4 array when (a) long range interactions are neglected, and (b) long range interactions are included. The positions of the peaks match the addition energy spectrum in Fig.~\ref{fig:eadd}, where elastic scattering of the electrons from the leads to the arrays is allowed. The width of the peaks is approximately $kT$.}\label{fig:condct}
\end{figure}

The conductance for a $3\times 4$ array is given in Fig.~\ref{fig:condct}. The positions of the peaks of the conductance are those values of the chemical potential that matches one of the addition energy in Fig.~\ref{fig:eadd}. A plot of the conductance for the case when the long-range interactions are neglected is shown for comparison. Without long-range interactions the conductance peaks cluster to the lower and upper Hubbard bands separated by the Mott gap. With long range interaction the peaks spread out more evenly and the Mott gap is less clearly visible. 

\section{Localization in the donor array}
We now investigate two types of localization in the array that has an important influence on transport. One is a consequence of the dominant long range electron-ion core attraction at low filling, and the other a result of the tunnel coupling oscillation arising from intervalley interference of the donor electron's wavefunction.

\subsection{Localization induced by  long range interactions}
\begin{figure}
\includegraphics[scale=0.55]{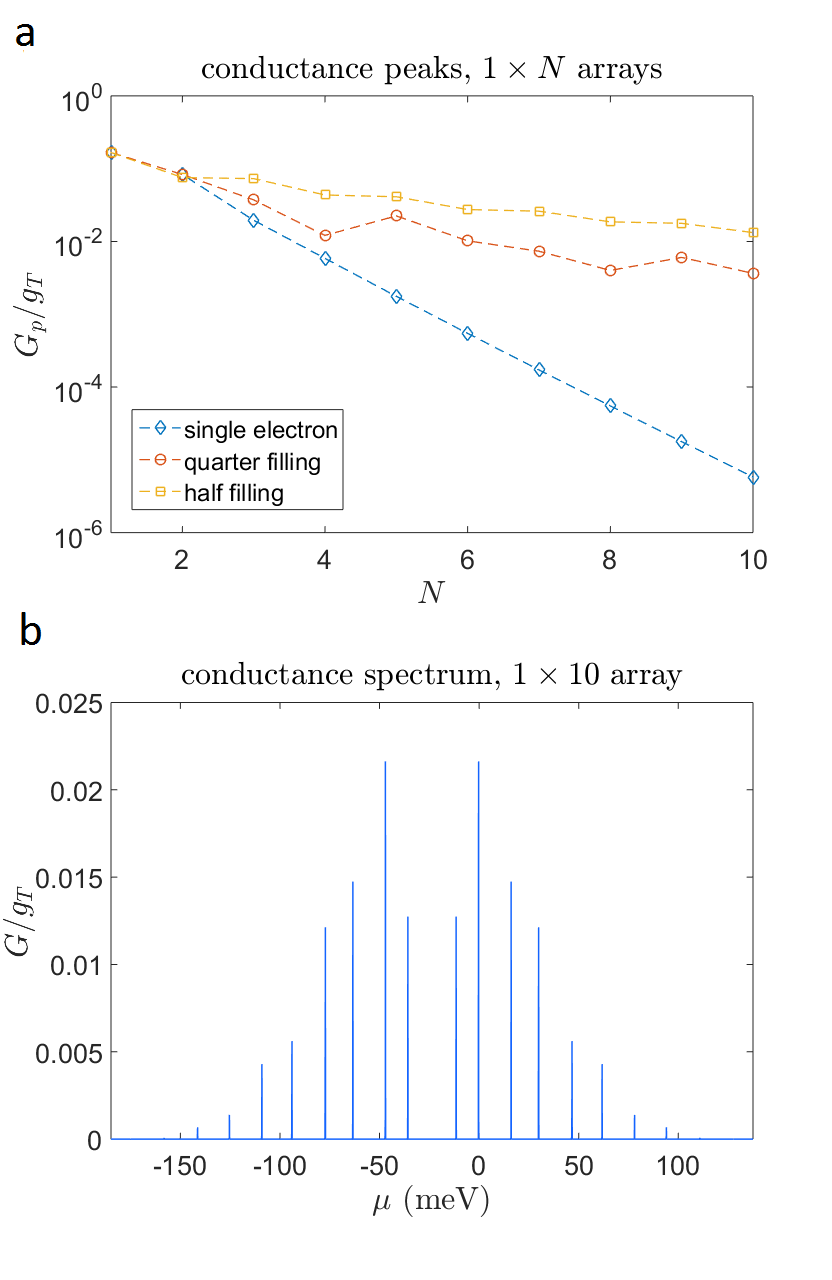}
\caption{(a) The scaling with the channel length in 1D of the conductance peaks at various filling: single electron (diamond), quarter filling (circle), and half filling (square); $T=100$ mK. (b) Conductance spectrum  of a 1D chain with 10 donors, which reveals how the peaks decrease when moving away from half filling.}\label{fig:gord}
\end{figure}

As illustrated in Fig.~\ref{fig:density}, the long range electron-ion core attraction leads to a localization of the electrons towards the center of the array at low filling. We show how this results in a faster decay of the conductance with system size. We compute the value of the conductance peak at single-electron filling (the first peak in the conductance spectrum), quarter- and half-filling, and study how these peaks scale with the number of donors in a 1D array, as shown in Fig.~\ref{fig:gord}(a).

The exponential decay at single electron filling is due to the localization of this electron's wavefunction at the center of the chain; the value of the conductance is proportional to the exponential tail at the edges. When the number of electrons is increased towards half filling, the electron-electron long range repulsion cancels the electron-ion core long range attraction, which leads to a reduction in the degree of localization. This explains why the conductance peak increases with increasing filling. This dependence of the conductance peak  on the filling is illustrated for a $1\times 10$ chain in Fig.~\ref{fig:gord}(b).

The resonant tunneling at half filling corresponds to the transition from $N-1$ to $N$ electrons in the array, where $N$ is the number of donors in the 1D chain. For our chosen nearest neighbor donor separation of $4.6$ nm, $U/t\approx 5$ which is quite large, the state at half filling can be thought of as having one localized electron occupying each site. The state with $N-1$ electrons has a single hole moving in a sea of localized electrons, and it is this hole that is responsible for transport. This hole spreads evenly through out the ordered array in a coherent superposition, thus the matrix element in Eq.~\eqref{eq:rate}  should scale as $1/N$. Our polynomial fit of the data for the conductance peak at half filling confirms this scaling.

\subsection{Localization induced by intervalley interference}\label{sec:condctdist}
\begin{figure}[t]
\includegraphics[scale=0.45]{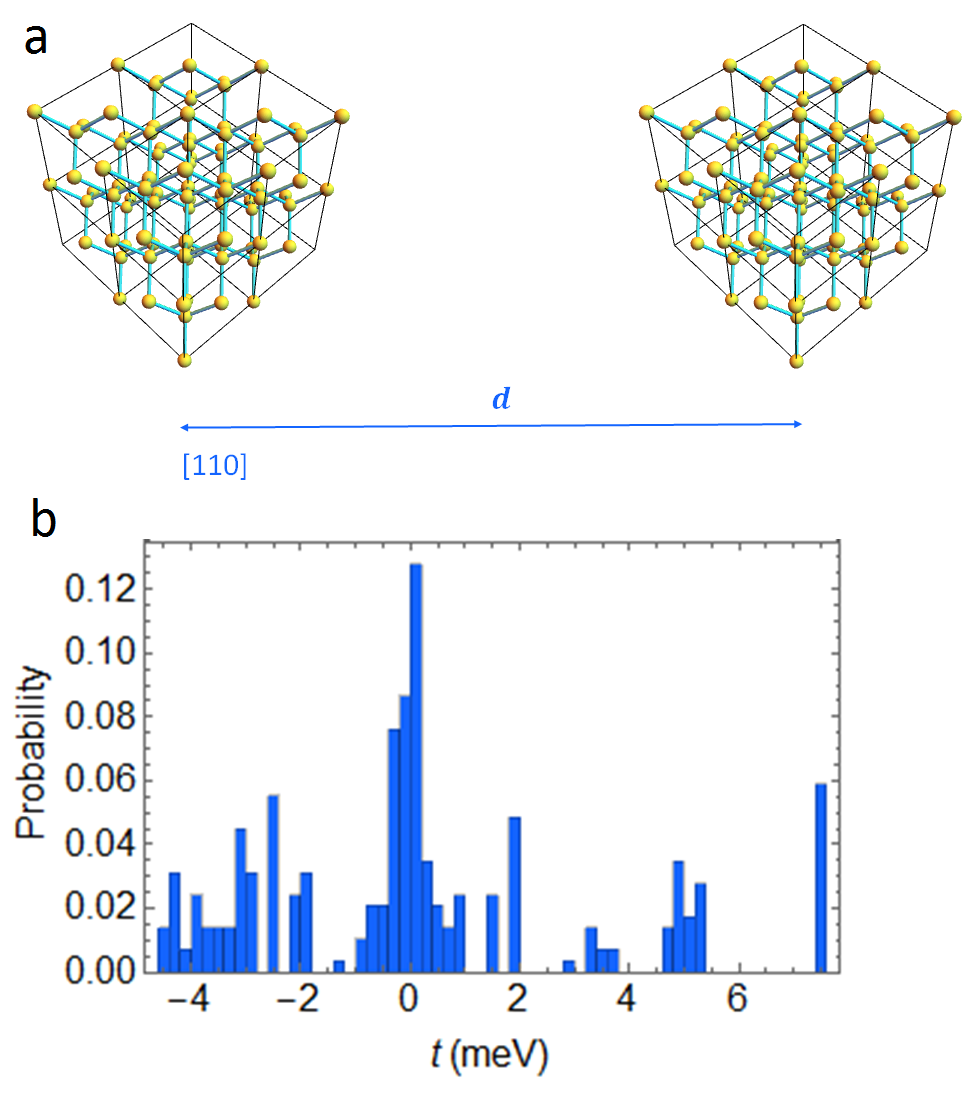}
\caption{(a) Two cubic unit cells of silicon whose centers are separated by 4.6 nm along the [110] direction. (b) Probability distribution of the tunnel coupling between two donors each occupying a substitutional site within its cubic unit cell. The donors are randomly placed at the sites according to a uniform distribution.}\label{fig:thist}
\end{figure}

We now discuss the realistic scenario when there is an uncertainty in the position of each donor in the array. STM-based lithography techniques allow nanometer precision. At first this seems to enable the fabrication of an almost perfectly ordered system where effect of disorder can be neglected. However, we see in this section that the intervalley interference of the $\text{1sA}_1$ wavefunction results in a large variation of the tunnel coupling even when the uncertainty in the donors' position is only on the order of $1$ nm. This disorder in the tunnel coupling results in so-called Lifshitz localization of the wavefunction \cite{B.I.Shklovskii788}, similar to Anderson localization arising from the disorder in single-site energy.   Therefore, it is important to ask whether we can still observe transport through delocalized states of the system as discussed in the previous section.

This problem was addressed recently in Ref.~\cite{dusko_mitigating_2016}; however, all the interactions, on-site and intersite, were neglected. Given the large magnitude of the interaction strength in the array, it is not clear whether this is justified. Our numerical calculation shows that for 1D chains the variation of the tunnel coupling can lead to a complete suppression of transport at low temperature, while the situation for 2D arrays is more promising owing to the larger number of possible paths available for conduction. Strong interactions in the system lead to a further enhancement of localization and hence a stronger suppression of transport in both 1D and 2D.

Figure \ref{fig:param} shows the oscillation in the tunnel coupling only for the case when the two donors are confined to a line along a crystal axis. In order to see the real extent of this oscillation we must allow the donors' positions to vary in three dimensions. For this we consider two donors within two cubic cells with sides equal to the lattice constant $a_{Si}=0.543$ nm, separated from center to center by $d=4.6$ nm along the [110] axis [see Fig.~\ref{fig:thist}(a)]. Each donor can randomly occupy any site within its cube according to a uniform distribution. We compute the tunnel coupling for all configurations and show its distribution in Fig.~\ref{fig:thist}(b). The values are spread out in the range between -4 and 8 meV. More importantly, there is around a 30$\%$ chance that the tunnel coupling almost vanishes, which may be a bottleneck for transport.

\begin{figure}[t]
\includegraphics[scale=0.65]{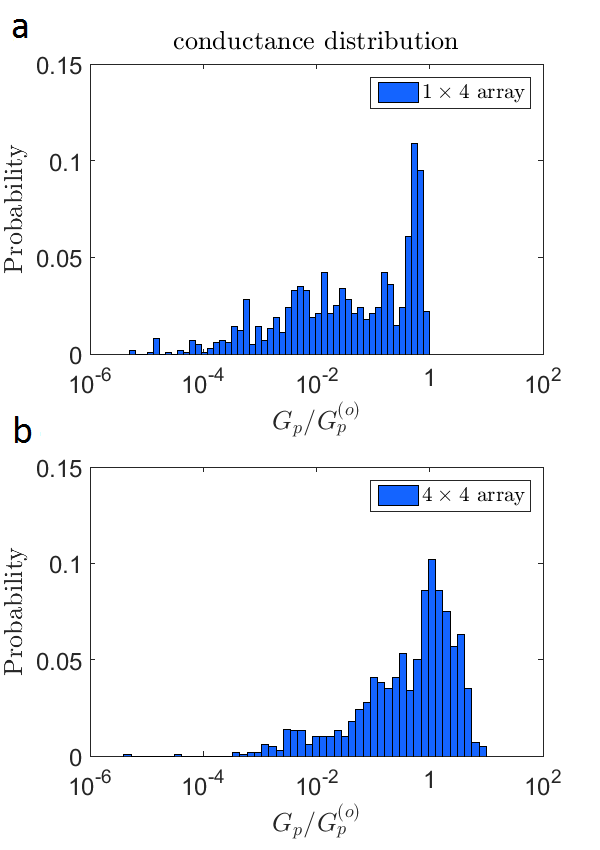}
\caption{Probability distribution of the conductance peak at half filling, scaled by its corresponding value in the perfectly ordered case, for a (a) $1\times4$ chain and (b) $4\times 4$ array. The arrays are oriented along the [110] axis with an average nearest neighbor separation of 4.6 nm. Each donor is randomly distributed within a cubic unit cell according to a uniform distribution. These results are obtained by exact diagonalization.}\label{fig:ghist}
\end{figure}

In order to see the effect of this disorder on the conductance, we generate many instances of an array where each donor can randomly occupy any site within its unit cube and compute the conductance peak at half filling at $T=100$ mK for each instance. We increase the sample size until we see little change in the shape of the conductance distribution. From a sample of 1000 instances we obtain the probability distribution in Fig.~\ref{fig:ghist} for a $1\times 4$ and a $4\times 4$ array. In order to speed up the calculation for the 2D array we keep only the contribution of the lowest energy level of each charge state in the conductance formula [see Eq.~\eqref{eq:gapprox} in the Appendix]; we have verified with smaller arrays that this approximation is sufficiently accurate at $T=100$ mK.

\begin{figure}[t]
\includegraphics[scale=0.6]{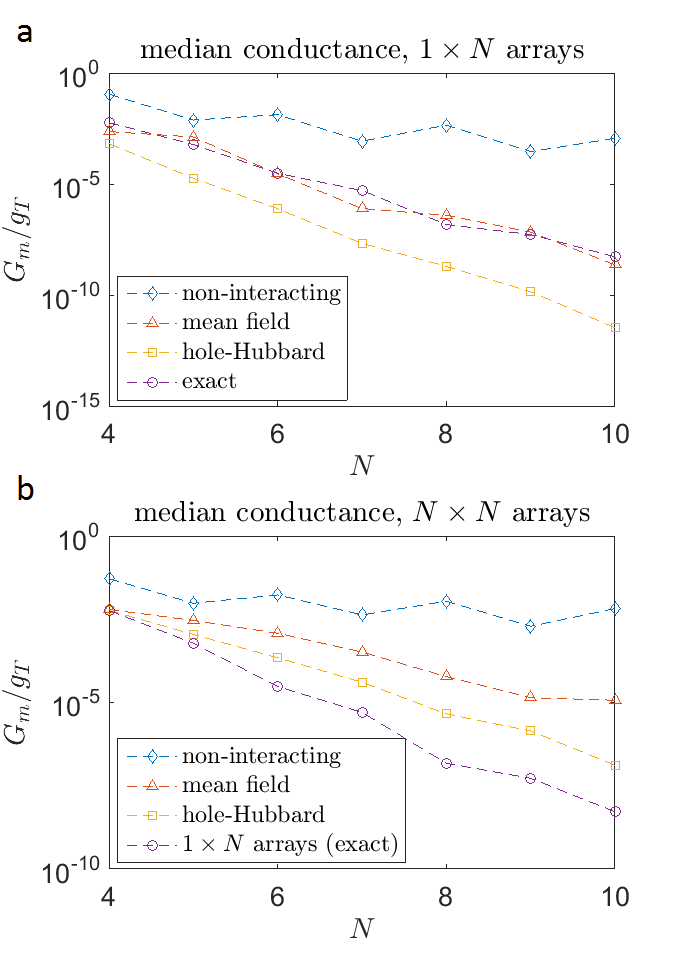}
\caption{(a) The scaling of the median value of the conductance distribution (at half filling) with system size for a (a) a 1D chain obtained by the noninteracting approximation (diamond), restricted Hartree-Fock mean field approximation (triangle), hole-Hubbard approximation (square) and exact diagonalization (circle), and (b) a 2D $N\times N$ array obtained by the noninteracting approximation (diamond), restricted Hartree-Fock mean field approximation (triangle) and hole-Hubbard approximation (square). The exact result for 1D (circle) is included in the bottom figure for a comparison with the mean field result for 2D.}\label{fig:gmedian}
\end{figure}

One sees from Fig.~\ref{fig:ghist} that localization in the 1D chain results in a significant probability that the conductance is suppressed by several orders of magnitude. This poses a challenge for the device fabrication process if the goal is to observe the resonant tunneling at low temperature in order to infer the underlying Hubbard physics in one dimension. Fortunately the suppression of transport is remarkably smaller for the $4\times 4$ array, owing to the larger number of paths available for conduction in two dimensions. Even if one path is blocked by a drop in the tunnel coupling, there are still paths that may support delocalized states for the charge excitation. 

As the number of sites along the channel length grows, the conductance suppression due to the variation in the tunnel coupling should be more prominent. We plot the median value of the conductance distribution at half filling for various channel lengths in Fig.~\ref{fig:gmedian} for 1D and 2D arrays. This median value is a relevant quantity as 50 $\%$ of the fabricated devices would yield a smaller conductance; thus it predicts whether resonant transport can be observed with high likelihood.

Exact diagonalization is not feasible for 2D arrays larger than $4\times 4$, so we use the Hartree-Fock mean field approximation \cite{xu_spin_2011}(see Appendix for more details). We also use another method which we call the hole-Hubbard approximation, which is efficient near half-filling. Since in our system $U$ is large compared with $t$ the half filled state can be thought of as having one localized electron occupying each site; when the number of electrons is $N_s-k$ with $k\ll N_s$ ($N_s$ is the number of sites), we can think of the transport mechanism as the hopping of the $k$ holes. This is the same idea behind the physical explanation of Nagaoka ferromagnetism \cite{tasaki_nagaokas_1998}. We write down an effective Hubbard Hamiltonian for these $k$ holes including the on-site and long range interaction between the holes (see the Appendix for more details). When $k$ is small we can save a lot of computational effort. Then the conductance peak of the transition between $N_s-k$ electrons and $N_s-k+1$ electrons can be approximated by the conductance peak of the transition between $k$ holes and $k-1$ holes. This approximation should be accurate when $U/t\gg 1$. One important difference between the mean field calculation and the hole-Hubbard approximation is that while the former neglects all the correlation between the electrons, the latter includes the correlation between the hopping holes, which are the quasiparticles responsible for the transport process.

For 1D arrays the mean-field calculation of the median conductance agrees very well with exact diagonalization [see Fig.~\ref{fig:gmedian}(a)]. The results in the noninteracting approximation, where both the on-site $U$ and intersite $V_0$ are set to zero, are also given, to illustrate the effect of the strong interactions in the system.  The fast exponential decrease for 1D arrays means that resonant transport at low temperature cannot be observed unless the number of sites is sufficiently small. The constant $g_T$ can  be estimated if one knows the tunneling rate $\Gamma$: Assuming a lead-donor coupling similar to the device of Ref.~\cite{fuechsle_single-atom_2012}, one can extract from the measurement results that $g_T\approx 10 \mu S$, yielding a $10 nA$ current at 1 meV bias. Our simulation predicts that this current is around 0.1 fA for a $1\times 10$ chain, which is too small to be detected.

The mean field calculation shows that the exponential decay for the conductance for 2D arrays is much slower. For $N=10$ the conductance is around five orders of magnitude larger than that in 1D. The current in a $10 \times 10$ array is in the region of 0.1 pA, which is detectable as demonstrated in a previous experiment with coupled donor transistors \cite{jehl_coupled_2015}. 

We see from Fig.~\ref{fig:gmedian}(a) that the hole-Hubbard approximation underestimates the conductance of 1D arrays. This is due to the assumption that at half filling each electron is perfectly localized at one site (there is no double occupancy), which is strictly valid only when $U/t\rightarrow \infty$. Therefore, the hole-Hubbard approximation overestimates the on-site interaction $U$ responsible for the  Mott insulating behavior of the electrons,  and hence underestimates the conductance.  For the 2D case we expect that the hole-Hubbard approximation also gives an underestimate; however, within this approximation, the conductance of the 2D array  is also around five orders of magnitude larger than that of the 1D array at $N=10$, which is consistent which the mean-field results.

A least squares fit of the mean field results shows that $G_m/g_T \sim 10^{-N}$ for 1D arrays and $G_m/g_T \sim 10^{-N/2}$ for 2D arrays. We remark that these scaling laws are applicable only for a finite array that is small enough so that its low lying energy spacing $\Delta E$ is large compared with the coupling $\Gamma$ to the leads. Nevertheless, the exponential decay in the presence of the large variation in the tunnel coupling is consistent with the universal scaling law for the  conductance in macroscopic strongly disordered systems \cite{abrahams_scaling_1979}.

\subsection{Discussions}
We end the paper with a discussion of the possible deviation of real measurements from our calculation and how to overcome some of the imperfections described above. First, one sees that a large portion of the upper Hubbard band in the conductance spectrum has energy above the conduction-band edge of silicon ($E_{ad}>0$), so a saturated plateau due to transport through the silicon conduction band should be observed in place of this part of the upper band. Second, the conduction in the upper Hubbard band is due to electrons tunneling through the $D^{-}$ state. This state has a larger orbital radius compared with $D^{0}$ \cite{larsen_stress_1981,tankasala_two-electron_2017} so the tunnel coupling should be larger and hence the amplitude of the peaks in the upper band should be larger than those in the lower band.

We see in Sec.~\ref{sec:gspectrum} that the long range interaction leads to a broadening of the Hubbard bands making it harder to identify the bands and the Mott gap. It also leads to a polynomial decay of the conductance with channel length in ordered arrays at low filling. This imperfection can be reduced by fabricating two parallel thin layers of saturated dose donors, one below and one above the array. These layers act as metallic plates that screen the long range interaction through image charges (see Appendix A). The fabrication of such a layer by STM lithography is demonstrated recently \cite{gramse_nondestructive_2017}.

In this paper we consider positional variation within only one unit cell. In a real sample the amount of positional variation is much larger than that. However, we repeat our simulation for a few different sizes of the array with the positional variation increased up to 2 nm, and we see very little change in the median value of the conductance distribution discussed in Sec.~\ref{sec:condctdist}. An explanation is that the sharp oscillation in the tunnel coupling leads to an already strong localization even with the minimal amount of positional variation, thus increasing the latter further does not lead to stronger localization and correspondingly smaller conductance.

In our simulation we find that the conductance distribution is sensitive to the tunneling coupling distribution, particularly to the existence of the central peak around zero in Fig.~\ref{fig:thist}. This peak appears in calculation using the ground state wavefuction obtained from the multivalley effective mass theory of Ref.~\cite{gamble_multivalley_2015}. When the simulation is repeated with another formula of the tunnel coupling given by the the Huckel's approximation in Ref.~\cite{dusko_mitigating_2016}, we find that the probability that the tunnel coupling drops to near zero is smaller and the median conductance of a $1\times 10$ array increases by around two orders of magnitude. We also note that there are theory models which predict that the tunnel coupling does not drop to near zero at all \cite{wellard_donor_2005}. The exact nature of the tunnel coupling oscillation is an open question and has not been addressed in experiments. It is likely that the median conductance we estimated in this paper is on the low side.

\section{Conclusions}
We have identified the appropriate extended Hubbard model that describes the physics of the low-lying states in an array of phosphorous donors in silicon. We show that the long-range interactions in the array have important effects far from  half-filling, for example the localization of the carriers towards the center of the array leading to a reduction of the conductance. These long range interactions should be taken into account in the analysis of experiments where the array is used as a quantum simulator of the Hubbard model. We also investigated the impact of the oscillation in the tunnel coupling due to the intervalley interference of the donor electron's wavefunction. This disorder is another mechanism for the localization of the many-body wavefunction, causing a sharp exponential decay of the conductance with system size in 1D. The situation is  more promising for a 2D array since there are a larger  number of paths for transport, and it is likely that the charge excitation responsible for conduction is delocalized, on the scale of the devices considered, along at least one of these paths. 

\begin{acknowledgements}
We would like to thank Benedict Murdin, Neil Curson, Alexander Koelker, Taylor Stock, Guy Matmon and Thornton Greenland for stimulating discussions and providing details on the STM-based fabrication process. We acknowledge financial support from the UK Engineering and Physical Sciences Research Council [COMPASSS/ADDRFSS, Grant No. EP/M009564/1] and EPSRC strategic equipment Grant No. EP/L02263X/1. The data underlying this work is available without restriction \cite{data}.
\end{acknowledgements}

\appendix
\section{Effect of image charge on single site energy}
When a donor is placed close to a metallic lead the induced charge in the lead is known to have a significant effect on the addition energy, as shown for a single donor in Ref.~\cite{calderon_heterointerface_2010}. Now we estimate how the induced charge affects the single site energy of the donor array. Consider a $N \times N$ square array (as in Fig.~\ref{fig:array}) whose bottom left donor  is separated from the left lead by $d_1$ and the right lead by $d_2$, with nearest neightbor donor separation $d$. Each donor's ion core induces charge in both the left and the right leads. Treating the leads as two infinite parallel grounded planes which are perpendicular to the plane of the donor array, the induced charge in the leads can be represented as a series of an infinite number of image charges  \cite{zahn_point_1976}. Choosing the coordinate origin at the site of the bottom left donor one can show that the potential caused by all the image charges of a point charge $q$ situated at $x_0, y_0$ is 
\begin{widetext}
\begin{align*}
V^{(\text{img})}(x,y)= -\frac{q}{4\pi\epsilon_0\epsilon_{Si}}\sum_{j=1}^{\infty} \bigg\{&\big[(x+x_0+2j L -2d_2)^2+(y-y_0)^2\big]^{-1/2} 
-\big[(x-x_0-2j L)^2+(y-y_0)^2 \big]^{-1/2} \\ \nonumber
+ &\big[(x+x_0-2 j L +2d_1)^2+(y-y_0)^2 \big]^{-1/2}
- \big[(x-x_0+2 j L)^2+(y-y_0)^2 \big]^{-1/2}\bigg\},
\end{align*}
\end{widetext}
where $L=d_1+d_2$ is the channel length. The electrostatic energy between a point charge $q$ in the array and a system of image charges is $(1/2) q V^{(\text{img})}_{\text{tot}}$ where $V^{(\text{img})}_{\text{tot}}$ is the total potential caused by all the image charges at the position of the point charge $q$.

The single site energy $\epsilon_i$ when there is only one electron in the array (the index $i$ is a pair of row and column indices) now has an additional contribution from the interaction with the image charges
\begin{align}
\epsilon_{i}=-E_B+\sum_{j}V_{ij}+\epsilon_{i}^{(\text{img})},
\end{align}
where $\epsilon_{i}^{(\text{img})}$ is computed as the sum of the interaction energy between the electron and all the image charges (including the image charges of the electron), and the interaction energy between the ion cores and the image charges of the electron. The electrostatic energy due to the interaction of the ion cores with each other and with their image charges do not depend on the electron filling and therefore can be excluded (we are concerned with the energy change when an electron is added to the array so we can set the energy at zero filling as the energy gauge).

\begin{figure}[t]
\includegraphics[scale=0.5]{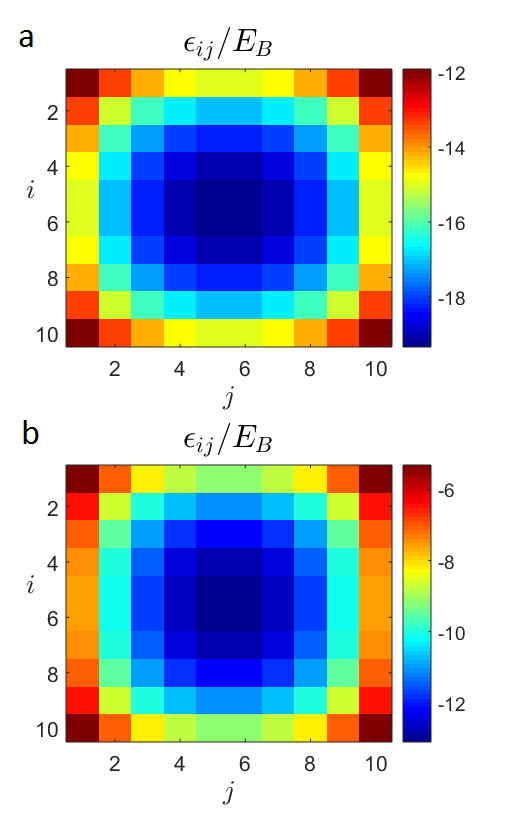}
\caption{A density plot of the single site energy, scaled by the binding energy of a neutral donor, in a $10 \times 10$ array on a square lattice when (a) the contribution from the image charge is excluded and (b) included ($d=4.6$ nm). The image charges lead to a large shift in the magnitude of the single site energy (note the difference in the two colobar scales) but there is little change in the energy variation.}\label{fig:eijdistimg}
\end{figure}

\begin{figure}[t]
\includegraphics[scale=0.5]{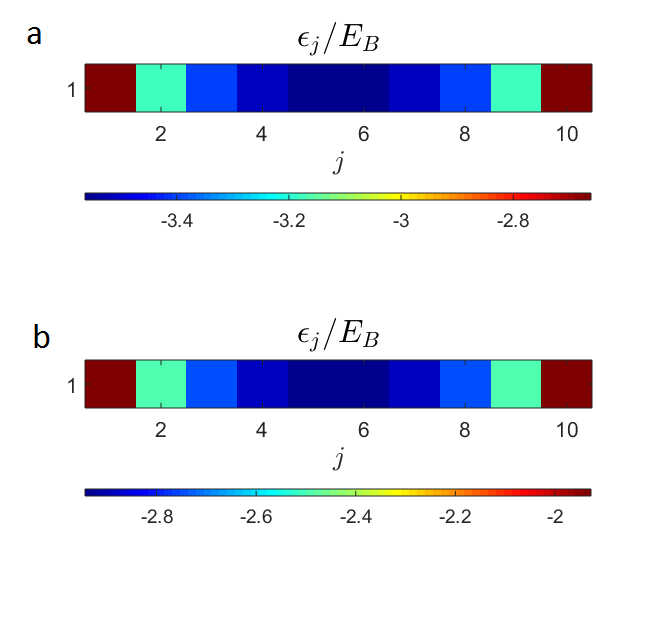}
\caption{A density plot of the single site energy in a $1 \times 10$ array when (a) the contribution from the image charge is excluded and (b) included ($d=4.6$ nm).}\label{fig:eijdist1D}
\end{figure}

Figure.~\ref{fig:eijdistimg} shows the single site energy distribution in a $10\times 10$ array with $d_1=10$nm, $d=4.6$nm and $d_2=9d+d_1$ (a symmetric setup) without and with the contribution from the image charges. Figure.~\ref{fig:eijdist1D} shows the same result for a 1D array. We see that the image charges lead to large shifts in the  magnitude of the single site energy, but it does not change significantly the variation in the energy from site to site. This is because the variation from the interaction of the electron with the far away image charges is small compared with the variation due to the interaction with the nearby ion cores. We expect the image charges do not lead to a significant change in the electron wavefunction at single electron filling. 

The problem becomes more complicated when there are more than one electron in the array since we have more image charges for electrons and these image charges are not static as the electrons hop around the array. One important effect is that the broadening of the Hubbard bands due to the long range interactions, as illustrated in Fig.\ref{fig:eadd}, is smaller because of the following: the lowest addition energy (for the transition from zero to one electron) can be approximated by $\min(\epsilon_i)$ and for the 2D array considered in Fig.~\ref{fig:eijdistimg} the image charges contribution shifts this value up by around $6 E_B$ from -19 $E_B$ to -13 $E_B$. Near half filing the array is almost neutral hence the contribution of the image charges should be small, thus there should be little change in the addition energy near half filling which is around -$E_B$ (one can work out this value by considering removing an electron from a half filled array). As a result the lower Hubbard band should be narrowed by 6 $E_B$. This narrowing of the band can be thought of as an effective reduction in the strength of the long range interaction. 

With STM lithography the source and the drain electrodes are not made as two metallic planes that are perpendicular to the plane of the donor array, but rather two sheets lying in the same plane. In this case the induced charge should be smaller compared with the induced charge in the setup treated above; and hence the effect of the image charges should be less noticeable.

\section{Hartree-Fock approximation}
In the Hartree-Fock approximation the extended Hubbard Hamiltonian is decomposed as
\begin{align}\label{eq:HubbardH}
&H_{array}=\sum_{i }\epsilon_{i} n_{i }-\sum_{\braket{i j} } \sum_{\sigma=\uparrow,\downarrow} t_{ij}\left(c^{\dagger}_{i \sigma} c_{j \sigma} +c^{\dagger}_{j \sigma} c_{i \sigma}\right)  \\ \nonumber
&+ U\sum_{i} n_{i\uparrow} \braket{n_{i \downarrow}}+U\sum_{i} \braket{n_{i\uparrow}} n_{i \downarrow}-U\sum_{i} \braket{n_{i\uparrow}} \braket{n_{i \downarrow}}\\
&+\sum_{i\neq j} W_{ij} n_i \braket{n_j}+\sum_{i\neq j} W_{ij} \braket{n_i} n_j-\sum_{i\neq j} W_{ij} \braket{n_i} \braket{n_j},
\end{align}
which is a single-particle Hamiltonian. For a 2D array each index $i$ is a pair of row index $j$ and column index $l$. At half filling the ground state is antiferromagnetic state so we start with the initial guess 
\begin{align}
&\braket{n_{jl\uparrow}}=1/2+m(-1)^{j+l},\\
&\braket{n_{jl\downarrow}}=1/2-m(-1)^{j+l},
\end{align}
where $m$ is a random number in the interval $[0,1/2]$. Single-particle states are then computed, which gives new values for $\braket{n_{jl\uparrow}}$ and $\braket{n_{jl\downarrow}}$. The computation is iterated until convergence. To avoid oscillations between two data points and speed up the convergence we use the mixture $[(1-c)\times \text{input}+c \times \text{output}]$ as the new input, where for each iteration $c$ is generated randomly in the interval $[0.1, 0.3]$.

When $kT$ is much smaller than the energy level splitting in the array we need to keep only the contribution from the lowest energy levels in the conductance formula of Eq.~\eqref{eq:condct}. One can show that within this approximation the conductance peak at half filling is given by
\begin{align}\label{eq:gapprox}
G_p\approx\frac{g_T}{6} \sum_{\sigma=\uparrow\downarrow}\frac{M^{(L)}_{\sigma}M^{(R)}_{\sigma}}{M^{(L)}_{\sigma}+M^{(R)}_{\sigma}};
\end{align}
where
\begin{align}
M^{(L)}_{\sigma}&=\sum_{j\in cL}  |\braket{\Psi^{n_{\sigma},n_{\bar{\sigma}}}_{\alpha=0}|c^{\dagger}_{j\sigma}|\Psi^{n_{\sigma}-1,n_{\bar{\sigma}}}_{\beta=0}}|^2, \\
M^{(R)}_{\sigma}&=\sum_{j\in cR}  |\braket{\Psi^{n_{\sigma},n_{\bar{\sigma}}}_{\alpha=0}|c^{\dagger}_{j\sigma}|\Psi^{n_{\sigma}-1,n_{\bar{\sigma}}}_{\beta=0}}|^2,
\end{align}
where $\alpha=\beta=0$ indicates the ground state and $n_{\sigma}=n_{\bar{\sigma}}=N^2/2$. We assume that the state $\Psi^{n_{\sigma}-1,n_{\bar{\sigma}}}_{\beta=0}$ can be approximated by removing the spin $\sigma$ particle with the highest filled single-particle orbital from the Hartree-Fock solution of $\Psi^{n_{\sigma},n_{\bar{\sigma}}}_{\alpha=0}$, then the matrix elements can be reduced to single-particle terms
\begin{align}
M^{(L)}_{\sigma}&=\sum_{j\in cL}  |\braket{\phi^{n_\sigma}_{\sigma}|c^{\dagger}_{j\sigma}|\text{vac}}|^2,\\
M^{(R)}_{\sigma}&=\sum_{j\in cR}  |\braket{\phi^{n_\sigma}_{\sigma}|c^{\dagger}_{j\sigma}|\text{vac}}|^2,
\end{align}
where $\ket{\text{vac}}$ is the vacuum and $\ket{\phi^{n_\sigma}_{\sigma}}$ is the highest filled single-particle orbital, which is readily available in the Hartree-Fock solution.

We used Matlab for our calculation. For arrays with a number of sites up to 12 we used the built in Matlab function for exact diagonalization. For the $4\times 4$ array studied in Sec.~\ref{sec:condctdist} a manually written Lanczos algorithm running on GPU is required. Arrays with number of sites larger than 16 are treated with Hartree-Fock approximation. We made these codes available on GitHub \cite{github}.

\section{The hole-Hubbard approximation}
As described in Sec.~\ref{sec:condctdist}, near half filling we can write down the following effective Hamiltonian for the hopping holes 
\begin{align}\label{eq:HubbardH}
H_{\text{hole}}=&\sum_{i }\epsilon_{i} n_{i }-\sum_{\braket{i j} } \sum_{\sigma=\uparrow,\downarrow} t_{ij}\left(c^{\dagger}_{i \sigma} c_{j \sigma} +c^{\dagger}_{j \sigma} c_{i \sigma}\right)  \\ \nonumber
 &+\sum_{i\neq j} W_{ij} n_i n_j,
\end{align}
where $\epsilon_i=E_B$ and $W_{ij}=V_0/|\mathbf{R}_i-\mathbf{R}_j|$. One major difference with the Hamiltonian for electrons is that there is no variation in the single-site energy due to long range Coulomb attraction. This is because in the picture of the hopping holes the sites without the hole have an electron and an ion core which forms a neutral center.

The conductance peak of the transition between $N_s-k$ electrons and $N_s-k+1$ electrons ($N_s$ is the number of sites) can then be approximated by the conductance peak of the transition between $k$ holes and $k-1$ holes, which can be estimated using Eq.~\eqref{eq:condct}.

\bibliography{donorarray}

\end{document}